\documentclass[11pt]{article}
\usepackage{amssymb}
\usepackage{latexsym}
\usepackage{amsmath}
\usepackage{graphicx}
\usepackage{bm}

\def\vp{\varphi}
\def\th{\theta}

\title{Dirac oscillator and nonrelativistic Snyder-de Sitter algebra }
\author{M.M. Stetsko\footnote{E-mail: mstetsko@gmail.com, mykola@ktf.franko.lviv.ua}\ \
\\
{\small Department of Theoretical Physics, Ivan Franko National University of Lviv,}\\
{\small 12 Drahomanov Str., Lviv, UA-79005, Ukraine}}
\begin{document}
\maketitle

\abstract{Three dimensional Dirac oscillator was considered in
deformed space obeyed to deformed commutation relations known as
Snyder-de Sitter algebra. Snyder-de Sitter commutation relations
gives rise to appearance minimal uncertainty in position as well
as in momentum. To derive energy spectrum and wavefunctions of the
Dirac oscillator supersymmetric quantum mechanics and shape
invariance technique was applied.}

\section{Introduction}
The Dirac oscillator represents an example of relativistic exactly
solvable quantum model. It was firstly proposed by It\^{o} and
collaborators to replace the momentum operator ${\bf P}$ in the
free particle's Dirac equation by combination ${\bf
P}-im\omega{\bf X}\beta$ where ${\bf X}$ was the position
operator, $m$ being the particle's mass and $\omega$ the
oscillator frequency. Then unusual accidental degeneracy of the
Dirac oscillator's spectrum was investigated by Cook
\cite{Cook_lett_NC_71}. Supersymmetric approach to the Dirac
oscillator was investigated in
\cite{Ui_PTP84,Balanntekin_AnnPhys85}. We note that the name Dirac
oscillator for this relativistic problem was given by Moshinsky
and Szczepaniak \cite{Moshinsky_JPA89} who rederived it and shown
that in nonrelativistic limit the relativistic hamiltonian becomes
a harmonic oscillator with a strong spin-orbit coupling term. The
last work renewed interest to the Dirac oscillator and it was
examined form different viewpoints, such as covariance properties
\cite{Moreno_JPA89}, complete energy spectrum and wavefunctions
\cite{BenitezPRL90}, Lie algebra symmetry \cite{Quesne_JPA90},
shift operators \cite{deLange_JPA91}, hidden supersymmetry
\cite{BenitezPRL90, Beckers_PRD90,Martinez_PRD91,Quesne_IJMPA91},
conformal invariance \cite{Martinez_JMP92}, completeness of
wavefunctions \cite{Szmytkowski_JPA01}, approach based on Clifford
algebra \cite{deLimaRodrigues_PLA08}. Some generalization of Dirac
oscillator was also considered \cite{Zarrinkamar_AnnPh10}.

The Dirac oscillator model was applied to problems of nuclear and
high energy physics. Relativistic many body systems with
interactions modelled by the Dirac oscillator hamiltonians with
applications to mesons and baryons was considered
\cite{Moshinsky_FoundPhys93}. Thermodynamics of Dirac oscillators
in $1+1$ spacetime was noted to be important in studies of
quark-gluon plasma \cite{Dominguez_EPL90}. It was also utilized
for developing of effective approach for description of
intermediate and short-range components of nucleon-nucleon
interaction \cite{Faessler_annPh05}. Dirac oscillator was used for
modelling photon-nucleus scattering \cite{Grineviciute_PRC09}.
Another area where the Dirac oscillator model was extensively
applied is quantum optics. Relation between the Dirac oscillator
and relativistic Jaynes-Cummings model was investigated
\cite{Rozmej_JPA99}. Mapping of the Dirac oscillator onto
Jaynes-Cummings model in case of different dimensions was examined
in \cite{Torres_AIPproc_2010}. In regard to the Jaynes-Cummings
model chirality quantum phase transition in $2+1$ dimensional
Dirac oscillator subjected to constant magnetic field was
investigated \cite{Bermudez_PRA08}. \textit{Zitterbewegung}
behaviour of the Dirac oscillator and possible realization of such
a system was considered in
\cite{Bermudez_PRA08_01,Romera_PRA_2011,Wang_EPJB_2012}. Several
attempts to get experimental realization of such a model were made
\cite{Longhi_OptLett10, Franco-Villafane}.

Here we consider the Dirac oscillator from a bit different point
of view, namely we solve the Dirac oscillator eigenavalue problem
in space with deformed Heisenberg algebra that lead to appearance
of minimal uncertainties in position and momentum. The interest to
the theories with deformed Heisenberg algebra was inspired by
investigations in string theory and independently by several
approaches to quantum gravity
\cite{GrossNPB88,Maggiore_PLB93,Witten_PhysToday96} where it was
suggested the existence of a finite lower bound for resolution of
length $\Delta X$, so called minimal length. Deformed commutation
relations that leads to existence of minimal uncertainty in
position and momentum was proposed  firstly by Kempf and
collaborators \cite{KMM_95} and then were investigated from
different viewpoints. We point out that only a few quantum
mechanical problems are solved exactly, that is to say the
harmonic oscillator in one \cite{KMM_95} and $D$ dimensions
\cite{Chang_PRD02}, the one- \cite{Noucier_JPA06} and
three-dimensional \cite{QuesneJPA05} Dirac oscillator and one
dimensional Coulomb-like problem \cite{Fityo_JPA06}.
Lorentz-covariant deformed algebra with minimal length was
proposed and $1+1$ dimensional Dirac oscillator problem was solved
\cite{Quesne_JPA06}. Minimal uncertainty for momentum can be
treated as a consequence of gravity induced decoherence
\cite{Kay}. Uncertainty relation that gives rise to appearance of
minimal momentum is also possible in theories with position
dependent effective mass \cite{Quesne_JPA_04_effmass}. We note
that deformed commutation relations with minimal length and
momentum were proposed even earlier in context of quantum group
theory \cite{Kempf_JMP94}. Later it was shown that similar
uncertainty principle with minimal length and momentum can be
obtained in a gedanken experiment of measuring of position in de
Sitter space \cite{Bambi_CQG}. Deformed algebra with minimal
length and momentum was also obtained in context of Triply Special
Relativity \cite{Kowalski-glikman_PRD04}. It should be noted that
basic principles of triply special relativity adopt three
fundamental constants and one of them can be identified with a
cosmological constant of de Sitter space. In case of deformed
algebra with minimal length and momentum only the harmonic
oscillator was examined
\cite{Quesne_JPA03,Quesne_JPA04,Mignemi_arxiv}.

Our paper is organized as follows. In the second section an
uncertainty relation obtained from deformed algebra is analyzed
then the Dirac oscillator oscillator is reviewed in given
representation. In the third section we obtain equations for small
and large components of a wavefunction and examine requirements
imposed on the wave function. In the forth section energy spectrum
of Dirac oscillator is obtained. In the fifth section
wavefunctions of the problem are derived. Finally, the sixth
section contains the conclusions.
\section{Dirac oscillator }
We consider stationary Dirac oscillator equation which can be
written in the form:
\begin{equation}\label{dirac_eq}
H\Psi=E\Psi, \quad H=\hat{{\bf\alpha}}({\bf P}-im\omega{\bf
X}\hat{\beta})+m\hat{\beta}
\end{equation}
where
\begin{eqnarray}
\hat{{\bf \alpha}}=
\begin{pmatrix}
0 & {\bf \sigma} \\
{\bf \sigma} & 0\\
\end{pmatrix}
\hat{\beta}=
\begin{pmatrix}
I & 0 \\
0 & -I\\
\end{pmatrix}\\
\end{eqnarray}
and $\sigma_i$ , $i=1,2,3$ are the Pauli matrices. We also put
$\hbar=c=1$. It is supposed that position $X_i$ and momentum $P_i$
operators in the equation \ref{dirac_eq} are obeyed to deformed
commutation relations which take form:
\begin{eqnarray}\label{algebra}
\begin{array}{l}
[X_i,P_j]=i\left(\delta_{ij}+\alpha X_iX_j+\beta
P_jP_i+\sqrt{\alpha\beta}(P_iX_j+X_jP_i)\right),\\
\\
{[X_i, X_j]=i\beta\varepsilon_{ijk}L_{k}}, \quad [P_i,
P_j]=i\alpha\varepsilon_{ijk}L_{k}.
\end{array}
\end{eqnarray}
Here $L_{k}$ are components of angular momentum operator and
parameters $\alpha$ and $\beta$ are supposed to be positive. We
also note, that there is summation over dummy indices. Components
of angular momentum operator are defined as follows:
\begin{equation}\label{ang_mom_def}
J_{ij}=\varepsilon_{ijk}L_{k}=\frac{1}{2}(X_iP_j+P_jX_i-X_jP_i-P_iX_j)
\end{equation}
Components of angular momentum operator fulfil the ordinary
commutation relations:
\begin{equation}
[L_i,X_j]=i\varepsilon_{ijk}X_k, \quad
[L_i,P_j]=i\varepsilon_{ijk}P_k.
\end{equation}
In the one-dimensional case the algebra (\ref{algebra}) takes
simpler form:
\begin{equation}\label{algebra_2}
[X,P]=i(1+\alpha X^2+\beta P^2+\sqrt{\alpha\beta}(PX+XP))
\end{equation}
We note that similar one-dimensional deformed algebra was examined
in the work \cite{Quesne_07sigma} but in their case instead of
factor $\sqrt{\alpha\beta}$ in the forth term in the right-hand
side an independent parameter $\kappa$ was used. It is easy to
show that the algebra (\ref{algebra_2}) gives rise to uncertainty
relation:
\begin{equation}\label{unceratinty}
\Delta X\Delta P\geqslant\frac{1}{2}|1+\gamma+\alpha(\Delta
X)^2+\beta(\Delta P)^2+\sqrt{\alpha\beta}\langle
\hat{X}\hat{P}+\hat{P}\hat{X}\rangle|
\end{equation}
where $\hat{X}=X-\langle X\rangle$, $\hat{P}=P-\langle P\rangle$
and $\gamma=(\sqrt{\alpha}\langle X\rangle+\sqrt{\beta}\langle
P\rangle)^2\geqslant 0$. From the inequality $|\langle
\hat{A}\hat{B}+\hat{B}\hat{A}\rangle|\leqslant 2\sqrt{\langle
A^2\rangle\langle B^2\rangle}$ which is valid for any two
operators $\hat{A}$ and $\hat{B}$ it follows that $|\langle
\hat{X}\hat{P}+\hat{P}\hat{X}\rangle|\leqslant 2\Delta X\Delta P$.
Since parameters $\alpha$ and $\beta$ are positive, it leads to
inequality $1+\gamma+\alpha(\Delta X)^2+\beta(\Delta P)^2>0$.
Using these remarks we can rewrite the uncertainty relation
(\ref{unceratinty}) in the form:
\begin{equation}\label{uncert_2}
\Delta X\Delta P\geqslant\frac{1}{2}(1+\gamma+\alpha(\Delta
X)^2+\beta(\Delta P)^2-2\sqrt{\alpha\beta}\Delta X\Delta P).
\end{equation}

The latter uncertainty relation brings minimal uncertainty in
position as well as in momentum:
\begin{equation}\label{min_uncert}
\Delta X\geqslant(\Delta
X)_{min}=\sqrt{\frac{\beta(1+\gamma)}{1+2\sqrt{\alpha\beta}}};
\quad \Delta P\geqslant(\Delta
P)_{min}=\sqrt{\frac{\alpha(1+\gamma)}{1+2\sqrt{\alpha\beta}}}
\end{equation}
It is important to emphasize that these minimal uncertainties do
not appear if parameters $\alpha$ and $\beta$ are negative. Having
done rescaling of uncertainties and parameters of deformation we
can represent uncertainty relation in the well known form obtained
by Kempf \cite{Kempf_JMP94}:
\begin{equation}\label{uncert_3}
\Delta\bar{X}\Delta\bar{P}\geqslant\frac{1}{2}(1+\bar{\alpha}(\Delta\bar{X})^2+\bar{\beta}(\Delta\bar{P})^2),
\end{equation}
where
\begin{equation}
\Delta\bar{X}=\sqrt{\frac{1+\sqrt{\alpha\beta}}{1+\gamma}}\Delta
X,\quad
\Delta\bar{P}=\sqrt{\frac{1+\sqrt{\alpha\beta}}{1+\gamma}}\Delta
P, \quad \bar{\alpha}=\frac{\alpha}{1+\sqrt{\alpha\beta}}, \quad
\bar{\beta}=\frac{\beta}{1+\sqrt{\alpha\beta}}. \nonumber
\end{equation}
It is no doubt that ``rescaled" uncertainty relation
(\ref{uncert_3}) leads to the same minimal uncertainties
(\ref{min_uncert}) as it should be.

In multidimensional case commutation relations (\ref{algebra})
brings to uncertainty relation:
\begin{equation}
\Delta X_i\Delta
P_j\geqslant\frac{1}{2}\left|\delta_{ij}+\gamma_{ij}+\alpha\langle\hat{X}_i\hat{X}_j\rangle+
\beta\langle\hat{P}_j\hat{P}_j\rangle+\sqrt{\alpha\beta}\langle\hat{P}_i\hat{X}_j+\hat{X}_j\hat{P}_i\rangle\right|
\end{equation}
where similarly as it was used in one dimensional case
$\hat{X}_i=X_i-\langle X_i\rangle$, $\hat{P}_i=P_i-\langle
P_i\rangle$ and $\gamma_{ij}=\alpha\langle X_i\rangle\langle
X_j\rangle+\beta\langle P_i\rangle\langle
P_j\rangle+2\sqrt{\alpha\beta}\langle P_i\rangle\langle
X_j\rangle$. It is easy to see that in case when $i=j$ the last
relation reduces to (\ref{uncert_2}) and as a consequence the
minimal uncertainties for position and momentum are the same as in
the one-dimensional case (\ref{min_uncert})

To solve the Dirac equation (\ref{dirac_eq}) representation of
operators $X_i$, $P_j$ that obeyed to commutation relations
(\ref{algebra}) should be defined. The algebra (\ref{algebra})
does not have position or momentum representations because of
noncommutativity of corresponding operators. To build up
representation for position and momentum operators (\ref{algebra})
it was proposed projective transformation \cite{Mignemi_arxiv}
which introduce a relation between the commutation relations
(\ref{algebra}) and Snyder algebra \cite{snyder}. As it was noted
such a transformation is nonsymplectic. The position and momentum
operators can represented as follows \cite{Mignemi_arxiv}:
\begin{eqnarray}
X_i=i\sqrt{1-\beta p^2}\frac{\partial}{\partial
p_i}+\lambda\sqrt{\frac{\beta}{\alpha}}\frac{p_i}{\sqrt{1-\beta
p^2}},\\
P_i=-i\sqrt{\frac{\alpha}{\beta}}\sqrt{1-\beta
p^2}\frac{\partial}{\partial
p_i}+(1-\lambda)\frac{p_i}{\sqrt{1-\beta p^2}}.
\end{eqnarray}
Here $p^2=p_kp_k$ and parameter $\lambda$ is arbitrary real. Since
$\alpha, \beta >0$ it leads to the restriction for the absolute
value of square of variable $p$: $\beta p^2<1$.

To provide hermicity of position and momentum operators scalar
product should be defined with a weight function. It can be
written in the form:
\begin{equation}\label{inner_product_general}
\langle\psi|\vp\rangle=\int\frac{d{\bf p}}{\sqrt{1-\beta
p^2}}\psi^*(\bf{p})\vp({\bf p})
\end{equation}
We note that according to abovementioned remark the domain of
integration is bounded by the sphere: $p^2\leqslant 1/\beta$. It
is worth emphasizing that the weight function does not depend on
the choice of parameter $\lambda$.

Components of the angular momentum operator defined by formula
(\ref{ang_mom_def}) are represented as follows:
\begin{equation}
J_{ij}=\varepsilon_{ijk}L_k=i\left(p_j\frac{\partial}{\partial
p_i}-p_i\frac{\partial}{\partial p_j}\right).
\end{equation}
So the components of angular momentum operator take the same form
as in momentum representation in ordinary quantum mechanics.

Wave function of the Dirac equation (\ref{dirac_eq}) can be
written as a two-component spinor $\psi=\begin{pmatrix}
\psi_1 \\\psi_2\\
\end{pmatrix}$ where functions $\psi_1$ and $\psi_2$ are called
large and small component respectively. The Dirac equation
(\ref{dirac_eq}) can be rewritten as a system of two coupled
equations:
\begin{eqnarray}
B^+\psi_2=(E-m)\psi_1,\label{eq_b_+}
\\ B^-\psi_1=(E+m)\psi_2\label{eq_b_-}.
\end{eqnarray}
where
\begin{equation}\label{operator_B_big}
B^{\pm}=(\bm{\sigma},{\bf P})\pm im\omega(\bm{\sigma},{\bf X})
\end{equation}
To get a factorized equation for the large component $\psi_1$ one
should apply the operator $B^+$ (\ref{operator_B_big}) to the
equation (\ref{eq_b_-}) and then in the right-hand sight of
obtained equation the action of the operator on the component
$\psi_2$ should be replaced by the right-hand side of the equation
(\ref{eq_b_+}). As a result we arrive at:
\begin{equation}
B^+B^-\psi_1=(E^2-m^2)\psi_1.
\end{equation}
Similarly for the small component $\psi_2$ we have:
\begin{equation}
B^-B^+\psi_2=(E^2-m^2)\psi_2.
\end{equation}
The representation of position and momentum operators $X_i$ and
$P_j$ allows one to get the explicit form for the operators
$B^{\pm}$:
\begin{equation}\label{oper_B_+}
B^+=\left[-i\left(\sqrt{\frac{\alpha}{\beta}}-im\omega\right)\sqrt{1-\beta
P^2}\left(\frac{\partial}{\partial p}+\frac{(\bm{\sigma}, {\bf
L})+2}{p}\right)+\left(1-\lambda+im\omega\lambda\sqrt{\frac{\beta}{\alpha}}\right)\frac{p}{\sqrt{1-\beta
p^2}}\right]\sigma_p
\end{equation}
\begin{equation}\label{oper_B_-}
B^-=\sigma_p\left[-i\left(\sqrt{\frac{\alpha}{\beta}}+im\omega\right)\sqrt{1-\beta
P^2}\left(\frac{\partial}{\partial p}-\frac{(\bm{\sigma}, {\bf
L})}{p}\right)+\left(1-\lambda-im\omega\lambda\sqrt{\frac{\beta}{\alpha}}\right)\frac{p}{\sqrt{1-\beta
p^2}}\right]
\end{equation}
where $\sigma_p=(\bm{\sigma},{\bf p})/p$

The equations (\ref{eq_b_+}), (\ref{eq_b_-}) and as a consequence
the operators $B^{\pm}$ would take simpler form if transformation
of large and small functions is performed:
\begin{equation}\label{subst}
\psi_i=\frac{1}{p}\vp_i
\end{equation}
After that transformation equations (\ref{eq_b_+}) and
(\ref{eq_b_-}) can be rewritten as follows:
\begin{eqnarray}
\tilde{\omega}b^+\sigma_p\vp_2=(E-m)\vp_1,\label{equ_1_b+}\\
\tilde{\omega}^*\sigma_pb^-\vp_1=(E+m)\vp_2.\label{equ_1_b-}
\end{eqnarray}
where
$\tilde{\omega}=\left(m\omega+i\sqrt{{\alpha}/{\beta}}\right)$ and
$\tilde{\omega}^*$ is complex conjugate. Operators $b^{\pm}$ are
obtained from relations (\ref{oper_B_+}), (\ref{oper_B_-}) and
(\ref{subst}). They take form:
\begin{equation}\label{op_b_+}
b^+=-\sqrt{1-\beta p^2}\frac{\partial}{\partial
p}-\frac{\sqrt{1-\beta p^2}}{p}((\bm{\sigma},{\bf
L})+1)+\eta\frac{p}{\sqrt{1-\beta p^2}}
\end{equation}
\begin{equation}\label{op_b_-}
b^-=\sqrt{1-\beta p^2}\frac{\partial}{\partial
p}-\frac{\sqrt{1-\beta p^2}}{p}((\bm{\sigma},{\bf
L})+1)+\eta^*\frac{p}{\sqrt{1-\beta p^2}},
\end{equation}
here
\begin{equation}
\eta=\frac{1-\lambda+im\omega\lambda\sqrt{\frac{\beta}{\alpha}}}{m\omega+i\sqrt{\frac{\alpha}{\beta}}}=
\frac{m\omega+i\left(m^2\omega^2\lambda\sqrt{\beta\over{\alpha}}-
\sqrt{\alpha\over{\beta}}(1-\lambda)\right)}{m^2\omega^2+\frac{\alpha}{\beta}}.\nonumber
\end{equation}
To simplify equations (\ref{equ_1_b-}) and (\ref{equ_1_b-}) one
can introduce function:
\begin{equation}
\tilde{\vp}_2=\sigma_p\vp_2
\end{equation}
As a result we arrive at
\begin{eqnarray}
\tilde{\omega}b^+\tilde{\vp}_2=(E-m)\vp_1,\label{equ_1x_b+}\\
\tilde{\omega}^*b^-\vp_1=(E+m)\tilde{\vp}_2.\label{equ_1x_b-}
\end{eqnarray}

\section{Superpartner hamiltonians and components of radial wave function}
Operators $b^{\pm}$ (\ref{op_b_+}) (\ref{op_b_-}) introduced in
the previous section commute with the total angular momentum ${\bf
J}={\bf L}+{\bf S}$ where ${\bf S}=\frac{1}{2}\bm{\sigma}$ as well
as with ${\bf L}^2$ and ${\bf S}^2$, so the solutions $\vp_1$ and
$\tilde{\vp}_2$ of equations (\ref{equ_1x_b+}) and
(\ref{equ_1x_b-}) can be taken in the form representing the fact
that they are eigenfunctions of operators ${\bf L}^2$, ${\bf
S}^2$, ${\bf J}^2$ and $J_z$ with corresponding eigenvalues
$l(l+1)$, 3/4, $j(j+1)$ and $m$ respectively.
\begin{equation}
\vp_1=\vp_1(p,s,j,m)=R_{1;s,j}(p)\mathcal{Y}_{s,j,m}(\th,\vp,\xi)
\end{equation}
\begin{equation}
\vp_2=\vp_2(p,s,j,m)=R_{2;s,j}(p)\mathcal{Y}_{s,j,m}(\th,\vp,\xi)
\end{equation}
where
\begin{equation}
\mathcal{Y}_{s,j,m}(\th,\vp,\xi)=\sum_{\sigma,\mu}\langle
j-s\mu,\frac{1}{2}\sigma|jm\rangle
Y_{j-s,m}(\th,\vp)\chi_{\sigma}(\xi)
\end{equation}
is a spin spherical harmonic \cite{Edmonds} and $R_{1;s,j}(p)$ and
$R_{2;s,j}(p)$ are radial wavefunctions. It should be noted that
$\chi_{\sigma}(\xi)$ denotes a spinor and $\sigma=\pm\frac{1}{2}$.

The main advantage of introduced function $\tilde{\vp}_2$ is
caused by the fact that it has the same spin-angular part as the
function $\vp_1$. Whereas for the function $\vp_2$ we have:
\begin{equation}
\vp_2=\sigma_p\tilde{\vp}_2=\tilde{R}_{2;s,j}(p)\sigma_p\mathcal{Y}_{s,j,m}(\th,\vp,\xi)=-\tilde{R}_{2;s,j}(p)\mathcal{Y}_{-s,j,m}(\th,\vp,\xi)
\end{equation}
Last relation can be written as follows:
\begin{equation}
\vp_2=\vp_{2;-s,j,m}(p,\th,\vp,\xi)=R_{2;-s,j}(p)\mathcal{Y}_{-s,j,m}(\th,\vp,\xi)
\end{equation}
and here $R_{2;-s,j}(p)=-\tilde{R}_{2;s,j}(p)$. We remark that
wavefunctions $\phi_1$ and $\tilde{\phi}_2$  are characterized by
the same value $l=j-s$.

To make equations (\ref{equ_1x_b+}) and (\ref{equ_1x_b-}) simpler
we consider relation:
\begin{equation}
((\bm{\sigma},{\bf L})+1) \mathcal{Y}_{s,j,m}(\th,\vp,\xi)=({\bf J}^2-{\bf L}^2-{\bf S}^2+1)\mathcal{Y}_{s,j,m}(\th,\vp,\xi)=s(2j+1)\mathcal{Y}_{s,j,m}(\th,\vp,\xi).
\end{equation}

Having used last equation one arrives at a system of coupled
equations for radial wavefunctions:
\begin{equation}\label{equ_3_b+}
\tilde{\omega}b_p^+\tilde{R}_2=(E-m)R_1,
\end{equation}
\begin{equation}\label{equ_3_b-}
\tilde{\omega}^*b_p^-R_1=(E+m)\tilde{R}_2.
\end{equation}
we use notation $b^{\pm}_p$ for the radial part of operators $b^{\pm}$ and they take form
\begin{equation}\label{b_p+}
b^+_p=-\sqrt{1-\beta p^2}\frac{\partial}{\partial p}-\frac{k}{p}\sqrt{1-\beta p^2}+\frac{\eta p}{\sqrt{1-\beta p^2}};
\end{equation}
\begin{equation}\label{b_p-}
b^-_p=\sqrt{1-\beta p^2}\frac{\partial}{\partial p}-\frac{k}{p}\sqrt{1-\beta p^2}+\frac{\eta^* p}{\sqrt{1-\beta p^2}};
\end{equation}
where $k=s(2j+1)$.

In radial momentum space the scalar product
(\ref{inner_product_general}) can be represented as follows:
\begin{equation}\label{inner_prod_radial}
\langle R|R'\rangle=\int^{1/\sqrt{\beta}}_0\frac{dp}{\sqrt{1-\beta p^2}}R^*(p)R'(p)
\end{equation}
It is easy to verify that with respect to the scalar product
(\ref{inner_prod_radial}) the operators $b_p^+$  (\ref{b_p+}) and
$b_p^-$ (\ref{b_p-}) are mutually hermitian conjugates.

From equations (\ref{equ_3_b+}) and (\ref{equ_3_b-}) we obtain:
\begin{eqnarray}
b_p^+b_p^-R_1=\frac{1}{|\tilde{\omega}|^2}(E^2-m^2)R_1;\label{equ_4_b+-}\\
b_p^-b_p^+\tilde{R}_2=\frac{1}{|\tilde{\omega}|^2}(E^2-m^2)\tilde{R}_2\label{equ_4_b-+}
\end{eqnarray}
The radial wavefunctions $R_1$ and $\tilde{R}_2$ can be treated as
eigenfunctions of two superpartner hamiltonians
\cite{Cooper_PhysRept95,Junker_96}.

We consider bound state problem so normalizability condition
should be imposed on the relativistic wavefunction
$\psi=\begin{pmatrix}\psi_1 \\\psi_2\\\end{pmatrix}$. It gives
rise to the following relation:
\begin{equation}\label{normal_wavefunct}
\int_0^{1/\sqrt{\beta}}\frac{dp}{\sqrt{1-\beta
p^2}}\left(|R_1|^2+|\tilde{R}_2|^2\right)=1.
\end{equation}
In the presence of deformed commutation relations additional
requirements are imposed on bound state wavefunctions. In case of
uncertainty principle with minimal length it is demanded that any
``physical" wavefunction belongs to the domain of operator
$\bf{P}$ it means that meanvalue of square of momentum operator is
finite. The deformed commutation relations (\ref{algebra}) impose
stricter requirements. To be acceptable a wavefunction should
belong to the domains of operators $\bf{P}$ and $\bf{X}$. As a
result it leads to finite meanvalues for square of both momentum
and position.

Let us suppose that in the right-hand side of equations
(\ref{equ_4_b+-}) and (\ref{equ_4_b-+}) we have eigenvalue
$E^2=m^2$, so the corresponding wavefunctions are necessarily the
solutions of equations:
\begin{eqnarray}
b_p^-R_{1;0}=0,\label{wavefunct_b_0-}\\
 b_p^+\tilde{R}_{2;0}=0.\label{wavefunct_b_0+}
\end{eqnarray}
 Having integrated equation (\ref{wavefunct_b_0-}) we obtain:
 \begin{equation}\label{wavfunct_r_backgr_zero}
 R_{1;0}=C_{1;0}p^k(1-\beta p^2)^{\frac{\tilde{\xi}}{2}+i\frac{\tilde{\zeta}}{2}}
 \end{equation}
 where $\tilde{\xi}=\frac{m\omega}{\alpha+\beta m^2\omega^2}$ and
 $\tilde{\zeta}=\frac{\sqrt{\alpha/\beta}(1-\lambda)-m^2\omega^2\lambda\sqrt{\beta/\alpha}}{\alpha+\beta
 m^2\omega^2}$ and $C_{1;0}$ is the normalization constant.

The normalization condition (\ref{normal_wavefunct}) implies that
integral from the square module of the function $R_{1;0}$ must be
finite:
\begin{equation}\label{funct_R^0_1}
\int^{1/\sqrt{\beta}}_0\frac{dp}{\sqrt{1-\beta
p^2}}\left|C_{1;0}\right|^2p^{2k}(1-\beta
p^2)^{\tilde{\xi}}=\left|C_{1;0}\right|^2\int^{1/\sqrt{\beta}}_0dp
p^{2k}(1-\beta p^2)^{\tilde{\xi}-\frac{1}{2}}<\infty
\end{equation}
For $p\rightarrow 0$ function $R_{1;0}$ behaves as $p^k$ and
boundary condition $R_{1;0}=0$ leads to the restriction $k>0$ and
this inequality is satisfied if $s=1/2$. When
$p\rightarrow\frac{1}{\sqrt{\beta}}$ convergence of the integral
(\ref{funct_R^0_1}) gives rise to the condition
$\tilde{\xi}-\frac{1}{2}>-1$ or equivalently
$\tilde{\xi}>-\frac{1}{2}$. But this inequality is always
fulfilled because the parameter $\tilde{\xi}$ is defined as
positive. So we conclude that wavefunction $R_{1;0}$ is
normalizable when $s=\frac{1}{2}$.

As it has been already mentioned additional ``physical" conditions
should be imposed on the wave function $R_{1;0}/p$. Meanvalues of
square of momentum and position operators must be finite:
\begin{equation}
\left\langle
\frac{R_{1;0}}{p}\Big|\hat{P}^2\Big|\frac{R_{1;0}}{p}\right\rangle<\infty,
\quad \left\langle
\frac{R_{1;0}}{p}\Big|\hat{X}^2\Big|\frac{R_{1;0}}{p}\right\rangle<\infty.
\end{equation}
Meanvalue for square of momentum can be represented in the form:
\begin{equation}\label{P_2_integral}
\left\langle
\frac{R_{1;0}}{p}\Big|\hat{P}^2\Big|\frac{R_{1;0}}{p}\right\rangle=\int^{1/\sqrt{\beta}}_0\frac{dp
p^2}{\sqrt{1-\beta
p^2}}\frac{R^*_{1;0}}{p}\hat{P}^2_p\frac{R_{1;0}}{p}<\infty,
\end{equation}
where:
\begin{eqnarray}\label{P_2_operator}
\hat{P}^2_p=-\frac{\alpha}{\beta}\left((1-\beta
p^2)\left(\frac{1}{p^2}\frac{\partial}{\partial
p}p^2\frac{\partial}{\partial p}-\frac{l(l+1)}{p^2}\right)-\beta
p\frac{\partial}{\partial p}\right)-\\\nonumber
2i\sqrt{\frac{\alpha}{\beta}}(1-\lambda)p\frac{\partial}{\partial
p}+(1-\lambda)\left((1-\lambda)-i\sqrt{\frac{\alpha}{\beta}}\right)\frac{p^2}{1-\beta
p^2}-3i\sqrt{\frac{\alpha}{\beta}}(1-\lambda)
\end{eqnarray}
is the ``radial" part of square of momentum operator. It should be
noted that all remarks concerning meanvalue of square of momentum
can be applied to the square of position operator because both of
them have similar structure.

Taking into account the explicit form for operator $\hat{P}^2_p$
(\ref{P_2_operator}) and using requirement (\ref{P_2_integral})
one can obtain following condition for integral:
\begin{equation}\label{int_converg}
\int^{1/\sqrt{\beta}}_0dp\quad p^{2k-2}(1-\beta
p^2)^{\tilde{\xi}-\frac{3}{2}}<\infty
\end{equation}
It is easy to convince oneself that convergence of the latter
integral in the vicinity of the point $p=0$ gives rise to the
condition $k>0$. From the other side convergence of the integral
(\ref{int_converg}) in the vicinity of the point $1/\sqrt{\beta}$
will be provided  if ${\tilde{\xi}-\frac{3}{2}}>-1$ from which we
obtain restriction on the parameters of oscillator if it is
supposed that parameters of deformation are held fixed:
\begin{equation}\label{cond_gr_state=0}
\frac{1}{\beta}(1-\sqrt{1-\alpha\beta})<m\omega<\frac{1}{\beta}(1+\sqrt{1-\alpha\beta})
\end{equation}
One can conclude that in order to obtain the eigenvalue $E^2=m^2$
in the equation (\ref{equ_4_b+-}) the condition $s=1/2$ should be
required. We note that in case of two-parametric deformed algebra
with minimal length eigenvalue $E^2=m^2$ exists also for positive
projection of spin $s=1/2$ but an additional demand for values $j$
must be satified \cite{QuesneJPA05}. The mentioned requirement
disappears in the limit case when one of those parameters
corresponding to our parameter $\beta$ is kept.

Having integrated equation (\ref{wavefunct_b_0+}) we obtain:
\begin{equation}\label{wavfunct_B_0+_integr}
\tilde{R}_{2;0}=C_{2;0}p^{-k}(1-\beta
p^2)^{-\frac{\tilde{\xi}}{2}+i\frac{\tilde{\zeta}}{2}}
\end{equation}
Again the boundary conditions are imposed on it. For the first we
require that $\tilde{R}_{2;0}\rightarrow 0$ when $p\rightarrow 0$.
The restriction $k<0$ or equivalently $s=-\frac{1}{2}$ follows
immediately from the last requirement. From the other side one
should demand $\tilde{R}_{2;0}\rightarrow 0$ when
$p\rightarrow\frac{1}{\sqrt{\beta}}$ but this requirement cannot
be fulfilled because $-\frac{\tilde{\xi}}{2}<0$. As a result, the
function $\tilde{R}_{2;0}$ is not normalizable. To have physically
acceptable function one should demand $\tilde{R}_{2;0}=0$ and
$R_{1;0}\neq 0$. It is worth mentioning that the same requirement
appears in case of two parametric deformed algebra with minimal
length \cite{QuesneJPA05}.  We also remark that the ground state
wavefunction $(R_{1;0}\neq 0, \tilde{R}_{2;0}=0)$ is compatible
with the positive eigenvalue $E=m$ whereas the negative one $E=-m$
will not be compatible with the system (\ref{equ_3_b+}) and
(\ref{equ_3_b-}).

\section{Spectrum of Dirac oscillator}
In this section we will obtain energy spectrum for the Dirac
oscillator. As it was shown in the previous section ground state
with energy $E^2=m^2$ exists only for positive projection of spin
($s=\frac{1}{2}$). In this section we will show that the ground
state with energy $E^2\neq m^2$ can take place for positive
($s=\frac{1}{2}$) as well as for negative ($s=-\frac{1}{2}$)
projection of spin. These two cases that correspond different
ground state energy are considered separately.
\subsection{Case of zero ground state energy}

As it has been already mentioned in the previous section that
whenever $s=\frac{1}{2}$ and the condition (\ref{cond_gr_state=0})
is fulfilled then equation (\ref{equ_4_b+-}) has acceptable
wavefunction corresponding to the ground state energy $E^2-m^2=0$.

To solve eigenvalue problem (\ref{equ_4_b+-}) SUSY QM procedure is
applied \cite{Cooper_PhysRept95,Junker_96}. Operator
$h=b^+_pb^-_p$ is supposed to be the first member of the SUSY QM
hierarchy
\begin{equation}
h_i=b^+_p(k_i,\eta_i)b^-_p(k_i,\eta_i)+\sum^i_{j=0}\varepsilon_j,
\quad i=0,1,2,\ldots
\end{equation}
Imposing shape invariance condition we obtain:
\begin{equation}\label{SI_condition}
b^-_p(k_i,\eta_i)b^+_p(k_i,\eta_i)=b^+_p(k_{i+1},\eta_{i+1})b^-_p(k_{i+1},\eta_{i+1})+\varepsilon_{i+1}
\end{equation}
In explicit form we write:
\begin{equation}\label{Im_eta_cond}
\eta_{i+1}-\eta^*_{i+1}=\eta_i-\eta_i^*
\end{equation}
\begin{equation}\label{k_cond}
k^2_{i+1}-k_{i+1}=k^2_i+k_i
\end{equation}
\begin{equation}\label{Re_eta_cond}
\frac{1}{\beta}|\eta_{i+1}|^2-\eta^*_{i+1}=\frac{1}{\beta}|\eta_{i}|^2+\eta_{i}
\end{equation}
\begin{equation}\label{epsilon_cond}
-k^2_{i+1}\beta-k_{i+1}(\eta_{i+1}+\eta^*_{i+1})-\frac{1}{\beta}|\eta_{i+1}|^2+\varepsilon_{i+1}=
-k^2_{i}\beta-k_{i}(\eta_{i}+\eta^*_{i})-\frac{1}{\beta}|\eta_{i}|^2
\end{equation}
In the following we use notations: ${\rm Re}\eta_i=\xi_i$ and
${\rm Im}\eta_i=\zeta_i$ Having solved the first three equations
we obtain:
\begin{equation}\label{iter_cond_e=0}
\zeta_{i}=\zeta, \quad \xi_{i}=\xi+\beta i, \quad k_i=k+i
\end{equation}
 we note that $\xi=\beta\tilde{\xi}$ and $\zeta=\beta\tilde{\zeta}$. It is easy to show that for obtained values $\eta_i$ and $k_i$
the hierarchy hamiltonians $h_i$ have physically acceptable
solutions $R_{1;0}(k_i,\eta_i,p)$ corresponding to the energies
$\Sigma^{i}_{j=0}\varepsilon_j$.

Having used the equation (\ref{epsilon_cond}) we arrive at
following equation for energy eigenvalues:
\begin{equation}\label{eigenvalues_general_e=0}
E^2_n-m^2=\left(m^2\omega^2+\frac{\alpha}{\beta}\right)\sum^n_{j=0}\varepsilon_j=4n\left(m^2\omega^2+\frac{\alpha}{\beta}\right)(\beta(n+k)+\xi)
\end{equation}
Since $k=s(2j+1)$ and
$\xi=\frac{m\omega}{m^2\omega^2+\alpha/\beta}$ the last relation
can be rewritten in the form:
\begin{equation}\label{spectrum_background_zero}
E^2_n-m^2=4n\left[m\omega+(m^2\omega^2\beta+\alpha)\left(n+j+\frac{1}{2}\right)\right]
\end{equation}
We note that in case $\alpha=0$  expression
(\ref{spectrum_background_zero}) is in agreement with
corresponding relation obtained in the work \cite{QuesneJPA05}
when one of their parameters of deformation is set to zero.

The principal quantum number $N=2n+l=2n+j-s$ can be introduced
instead of $n$. Then the relation (\ref{spectrum_background_zero})
can be represented as follows:
\begin{equation}
E^2_n-m^2=2\left(N-j+\frac{1}{2}\right)
\left[m\omega+\frac{1}{2}(m^2\omega^2\beta+\alpha)\left(N+j+\frac{3}{2}\right)\right].
\end{equation}
\subsection{Nonzero ground state energy}
Now we suppose that in the right-hand side of the equations
(\ref{equ_4_b+-})  and (\ref{equ_4_b-+}) we have $E^2-m^2\neq 0$.
It will be shown that in this case the ground state exists for the
following hamiltonian:
 \begin{equation}\label{gr_state_0}
h_0=b^+_p(k,\eta)b^-_p(k,\eta)=-\left(\sqrt{1-\beta p^2}\frac{\partial}{\partial p}\right)^2+(\eta-\eta^*)p\frac{\partial}{\partial p}+\frac{k^2-k}{p^2}+
\frac{\frac{1}{\beta}|\eta|^2-\eta^*}{1-\beta p^2}-k(\eta+\eta^*)-k^2\beta-\frac{1}{\beta}|\eta|^2
\end{equation}
In order to obtain ground state energy one should re-factorize hamiltonian $h_0$. It can be represented as follows:
\begin{equation}\label{gr_state_1}
h_0=b^+_p(k',\eta')b^-_p(k',\eta')+\varepsilon',
\end{equation}
where $k'$ and $\eta'$ are new parameters in operators
(\ref{b_p+}) and (\ref{b_p-}) and $\varepsilon'$ defines the
ground state energy. From equations (\ref{gr_state_0}) and
(\ref{gr_state_1}) it follows:
\begin{equation}\label{gr_st_im_eta}
\eta'-\eta'^*=\eta-\eta^*
\end{equation}
\begin{equation}\label{gr_st_k}
k'^2-k'=k^2-k
\end{equation}
\begin{equation}\label{gr_st_re_eta}
\frac{1}{\beta}|\eta'|^2-\eta'=\frac{1}{\beta}|\eta|^2-\eta
\end{equation}
\begin{equation}\label{gr_st_epsilon}
-k'(\eta'+\eta'^*)-\beta k'^2-\frac{1}{\beta}|\eta'|^2+\varepsilon=-k(\eta+\eta^*)-\beta k^2-\frac{1}{\beta}|\eta|^2
\end{equation}
Solving the equations (\ref{gr_st_im_eta})-(\ref{gr_st_re_eta})
we arrive at the relations:
\begin{eqnarray}
k'_1=k, \quad k'_2=1-k;\label{cond_gr_st_k}\\
\zeta'=\zeta,\quad \xi'_1=\xi, \quad
\xi'_2=\beta-\xi.\label{cond_gr_st_eta}
\end{eqnarray}
Since conditions for parameters $k'$ (\ref{cond_gr_st_k}) and
$\eta'$ (\ref{cond_gr_st_eta}) are obtained independently then one
can combine different $k'$ and $\eta'$ to investigate whether
obtained wavefunctions will be physically acceptable.

For the first we consider the case $k'=k$ and $\xi'=\xi$. It
follows immediately that $\varepsilon=0$. So the latter
combination should be left out.

Then if we suppose that $k'=k$ and $\xi'=\beta-\xi$ equation
(\ref{wavefunct_b_0-}) gives us corresponding wavefunction
$R_{1;0}=C_{1;0}p^k(1-\beta
p^2)^{\frac{\beta-\xi}{2\beta}+i\frac{\zeta}{2\beta}}$. The first
requirements imposed on this function are boundary conditions. To
provide the condition $R_{1;0}=0$ at the boundaries one should
demand that $k>0$ and $\beta-\xi>0$. As a consequence condition
$k>0$ leads to requirement $s=\frac{1}{2}$ whereas the demand
$\beta-\xi>0$ gives rise to
$m\omega\in(0,\frac{1-\sqrt{1-4\alpha\beta}}{2\beta})$ or
$m\omega\in(\frac{1+\sqrt{1-4\alpha\beta}}{2\beta}, \infty)$. If
$4\alpha\beta>1$ the last condition can be satisfied for arbitrary
$m\omega$. To make obtained wave function physically acceptable it
must fulfil normalizability condition (\ref{funct_R^0_1}) and even
stronger requirement (\ref{P_2_integral}). From the last
requirement it follows that $-\frac{\xi}{\beta}+\frac{1}{2}>0$ and
as a result it gives rise to the restrictions for the product
$m\omega$:
\begin{equation}\label{cond_nonzero_gr_state}
m\omega\in\left(0,\frac{1-\sqrt{1-\alpha\beta}}{\beta}\right)\bigcup\left(\frac{1+\sqrt{1-\alpha\beta}}{\beta},
\infty\right).
\end{equation}
One can see that obtained restrictions for the product $m\omega$
are opposite to (\ref{cond_gr_state=0}). As a conclusion, if the
relation (\ref{cond_gr_state=0}) is fulfilled then the ground
state has zero energy. If the condition (\ref{cond_gr_state=0}) is
broken the ground state with nonzero energy appears.

From relation (\ref{gr_st_epsilon}) we obtain
\begin{equation}\label{gr_st_energ_s=1/2}
\varepsilon=(\beta-2\xi)(1+2k).
\end{equation}
It is easy to verify that obtained ground state energy is
positive.

To find other eigenvalues of the hamiltonian $h_0$ one should
substitute $\xi'$ instead of $\xi$ into
(\ref{eigenvalues_general_e=0}) and take into account relation
(\ref{gr_st_energ_s=1/2}). After necessary transformations we
arrive at:
\begin{equation}\label{spectrum_s=1/2_E}
E^2_n-m^2=4(n+j+1)\left[-m\omega+(m^2\omega^2\beta+\alpha)\left(n+\frac{1}{2}\right)\right]
\end{equation}
Similarly as in previous case the obtained relation can be
represented in terms of principal quantum number
\begin{equation}
E^2_n-m^2=2\left(N+j+\frac{5}{2}\right)\left[-m\omega+\frac{1}{2}
(m^2\omega^2\beta+\alpha)\left(N-j+\frac{3}{2}\right)\right].
\end{equation}

From relations (\ref{cond_gr_st_k}) and (\ref{cond_gr_st_eta}) it
follows that ground states with nonzero energy other combinations
of $k'$ and $\eta'$ are also possible. We consider combination
$k'=1-k$ and $\xi'=\xi$. Then the ground state wavefunction takes
form: $R_{1;0}=C_{1;0}p^{1-k}(1-\beta
p^2)^{\frac{\xi}{2\beta}+i\frac{\zeta}{2\beta}}$. One of boundary
conditions leads to restriction $k<0$ which can be satisfied if
$s=-\frac{1}{2}$ and then $k'=j+\frac{3}{2}$. Since $\xi>0$ the
second boundary condition is satisfied immediately. It is easy to
persuade oneself that obtained wavefunction is normalizable.
Similarly as in the previous cases to make the obtained
wavefunction physically acceptable we should impose condition
(\ref{P_2_integral}) on it. Having used formula
(\ref{gr_st_epsilon}) we obtain following relation for the ground
state energy:
\begin{equation}
\varepsilon=(\beta+2\xi)(1-2k)
\end{equation}
One can see that ground state energy is also positive. Using the
same procedure as in case with positive $k$ on can obtain energy
spectrum:
\begin{equation}
E^2_n-m^2=4(n+j+1)\left[m\omega+(m^2\omega^2\beta+\alpha)\left(n+\frac{1}{2}\right)\right]
\end{equation}
Having introduced the principal quantum number we rewrite the last
relation in the form:
\begin{equation}
E^2_n-m^2=2\left(N+j+\frac{3}{2}\right)\left[m\omega+\frac{1}{2}
(m^2\omega^2\beta+\alpha)\left(N-j+\frac{1}{2}\right)\right].
\end{equation}
In the end we can choose combination $k'=1-k$ and
$\xi'=\beta-\xi$. This variant leads to the wave function
$R_{1;0}=C_{1;0}p^{1-k}(1-\beta
p^2)^{\frac{\beta-\xi}{2\beta}+i\frac{\zeta}{2\beta}}$. One of the
boundary conditions gives rise to the demand $k<0$ or equivalently
$k'=j+\frac{3}{2}$. Another boundary condition leads to inequality
$\beta-\xi>0$ but as we already know normalizability condition and
boundness of the square of momentum operator should be satisfied
and as a consequence all these demands lead to the condition
(\ref{cond_nonzero_gr_state}).

For the ground state we obtain:
\begin{equation}\label{gr_st_4}
\varepsilon=4(\beta(1-k)-\xi).
\end{equation}
It can be shown that the ground state energy (\ref{gr_st_4}) is
positive if $4\alpha\beta>1$. The same procedure leads us to the
following expression for the spectrum:
\begin{equation}
E^2_n-m^2=4(n+1)\left(-m\omega+(m^2\omega^2\beta+\alpha)\left(n+j+\frac{3}{2}\right)\right)
\end{equation}
Again we rewrite obtained formula replacing the quantum number $n$
by the principal quantum number $N$:
\begin{equation}
E^2_n-m^2=2\left(N-j+\frac{3}{2}\right)\left[-m\omega+(m^2\omega^2\beta+\alpha)
\left(N+j+\frac{5}{2}\right)\right].
\end{equation}

We note that this case does not have ``classical" limit or in
other words when parameters of deformation
$\alpha,\beta\rightarrow 0$ obtained spectrum does not reduce to
any solution of ordinary quantum mechanics \cite{Moshinsky_JPA89}.
Similar situation appears in the case of deformed algebra with
minimal length \cite{QuesneJPA05}.
\section{Radial momentum wavefunctions of Dirac oscillator}
In the previous section the ground state wavefunctions of
hamiltonian $h=b^+_pb^-_p$ have been derived. As it was shown only
the large component of wavefunction can be obeyed to all imposed
requirements. In this section we calculate the remaining large and
small components of radial momentum wavefunction
\subsection{Zero ground state energy}
The large component of radial momentum wavefunction for excited
states can be calculated with help of well-known SUSY QM and SI
technique \cite{Cooper_PhysRept95,Junker_96}. As it is known the
wave functions of the excited states are derived form the ground
state wavefunction with help of recursive procedure which is based
on the relation:
\begin{equation}\label{iter_proc_wf}
R_{1;n} (p;k,\xi)=\frac{1}{\sqrt{e_n-e_0}}b^+_p(k,\xi)R_{1;n-1}(p;k_1,\xi_1).
\end{equation}
Where we used notation $e_i=(E^2_i-m^2)/|\tilde{\omega}|^2$ for
simplicity. According to the conditions (\ref{iter_cond_e=0}) and
(\ref{eigenvalues_general_e=0}) we should impose $e_0=0$,
$k_1=k+1$ and $\xi_1=\xi+\beta$.

Having substituted the explicit form of operator $b^+_p$ into the
relation (\ref{iter_proc_wf}) we arrive at the equation:
\begin{equation}
R_{1;n} (p;k,\xi)=\frac{1}{\sqrt{e_n}}\left(-\sqrt{1-\beta p^2}\frac{\partial}{\partial p}-\frac{k}{p}\sqrt{1-\beta p^2}+\frac{(\xi+i\zeta)p}{\sqrt{1-\beta p^2}}\right)R_{1;n-1}(p;k+1,\xi+\beta).
\end{equation}
Given recursion procedure leads to consequence that the large
component of radial wavefunction takes form:
\begin{equation}\label{wf_e=0}
R_{1;n}(p;k,\xi)=C_{1;n}(k,\xi)p^{b+\frac{1}{2}}(1-\beta p^2)^{\frac{a}{2}+\frac{1}{4}-i\frac{\zeta}{2\beta}}P^{(a,b)}_n(z),
\end{equation}
where $C_{1;n}(k,\xi)$ and $P^{(a,b)}_n(z)$ are a normalization
factor a Jacobi polynomial respectively. Here we also denoted:
\begin{equation}
a=\frac{\xi}{\beta}-\frac{1}{2}, \quad b=k-\frac{1}{2}, \quad
z=2\beta p^2-1 \quad (-1<z<1).
\end{equation}

It was argued in the previous section that the small component of
the ground state radial wavefunction vanishes
$\tilde{R}_{2;0}(p;k,\xi)=0$. For excited states small component
can be found by using relation (\ref{equ_3_b-}):
\begin{equation}\label{wf_tilde_R_2}
\tilde{R}_{2;n}(p;k,\xi)=\frac{\tilde{\omega}^*}{E_n+m}b^-_p(k,\xi)R_{1;n}(p;k,\xi)
\end{equation}
Taking into account explicit expressions for the operator
$b^-(k,\xi)$ (\ref{b_p-}) and wave function (\ref{wf_e=0}) one can
rewrite the last relation in form:
\begin{eqnarray}\label{wf_e=0_2}
 \tilde{R}_{2;n}(p;k,\xi)=\frac{\tilde{\omega}^*C_{1;n}(k,\xi)}{E_n+m}\sqrt{1-\beta p^2}\left(\frac{\partial}
{\partial p}-\frac{k}{p}+\frac{\eta^* p}{1-\beta
p^2}\right)p^{b+\frac{1}{2}}(1-\beta
p^2)^{\frac{a}{2}+\frac{1}{4}-i\frac{\zeta}{2\beta}}P^{(a,b)}_n(z)\\\nonumber
=\frac{\tilde{\omega}^*C_{1;n}}{E_n+m}\sqrt{1-\beta
p^2}\left(\frac{\partial} {\partial
p}-\frac{b+\frac{1}{2}}{p}+\frac{\left(\beta\left(a+\frac{1}{2}\right)-i\zeta\right)
p}{1-\beta p^2}\right)p^{b+\frac{1}{2}}(1-\beta
p^2)^{\frac{1}{2}\left(a+\frac{1}{2}\right)-i\frac{\zeta}{2\beta}}P^{(a,b)}_n(z)\\\nonumber
=\frac{2\beta
\tilde{\omega}^*C_{1;n}(k,\xi)(n+a+b+1)}{E_n+m}p^{b+\frac{3}{2}}(1-\beta
p^2)^{\frac{1}{2}\left(a+\frac{3}{2}\right)-i\frac{\zeta}{2\beta}}P^{(a+1,b+1)}_{n-1}(z).
\end{eqnarray}
It should be noted that a formula of differentiation of the Jacobi
polynomials was used here \cite{Bateman_1953,Abramowitz}. In the
previous section it was stated that wavefunction
$(R_{1;0}(p;k,\xi)\neq 0,\tilde{R}_{2;0}(p;k,\xi)=0)$ is the
physically acceptable solution of the system of equations
(\ref{equ_3_b+}) and (\ref{equ_3_b-}) only for $ E^2_0=m^2$. At
the same time for excited states: $n=1,2,\ldots$ the solution of
this system of equations is given by
$(R_{1;n}(p;k,\xi),\tilde{R}_{2;n}(p;k,\xi))$. It is necessary to
verify whether these function are physically acceptable or not. It
is easy to persuade oneself that the Jacobi polynomials in
(\ref{wf_e=0}) and (\ref{wf_e=0_2}) do not spoil the convergence
of integral (\ref{normal_wavefunct}) and also meanvalues for
square of momentum and position operators would be finite
similarly as it was given by the condition (\ref{P_2_integral})
for the ground state wavefunction. Finally, the normalization
factor $C_{1;n}$ can be found from the normalization condition
(\ref{normal_wavefunct}):
\begin{equation}
C_{1;n}=\left(\beta^{b+1}(2n+a+b+1)\frac{n!\Gamma(n+a+b+1)}{\Gamma(n+a+1)\Gamma(n+b+1)}\frac{E_n+m}{E_n}\right)^{\frac{1}{2}}
\end{equation}
\subsection{Nonzero groundstate energy}
To find wavefunctions of excited states in remaining cases one
should follow the approach used in the previous section.
Parameters $k$ and $\xi$ in the iteration equation
(\ref{iter_proc_wf}) should be replaced by $k'$ and $\xi'$
correspondingly. It worth noting that at the same time parameters
$k$ and $\xi$ in the equation (\ref{wf_tilde_R_2}) remain
unchanged. As a consequence we can state that equation
(\ref{wf_e=0_2}) remains valid if parameters $a$ and $b$ are
replaced by a new one.

In the case $k'=k$ that corresponds $s=\frac{1}{2}$ and
$\xi'=\beta-\xi$ we obtain:
\begin{equation}
R_{1;n}(p;k,\xi)=C_{1;n}p^{b+\frac{1}{2}}(1-\beta
p^2)^{\frac{1}{2}\left(a+\frac{1}{2}\right)-i\frac{\zeta}{2\beta}}P^{(a,b)}_n(z)
\end{equation}
where $a=\frac{1}{2}-\frac{\xi}{\beta}$ and $b=k-\frac{1}{2}$.

The relation (\ref{wf_tilde_R_2}) gives rise to:
\begin{eqnarray}
\tilde{R}_{2;n}(p;k,\xi)=\frac{\tilde{\omega}^*C_{1;n}}{E_n+m}\sqrt{1-\beta
p^2}\left(\frac{\partial} {\partial p}-\frac{k}{p}+\frac{\eta^*
p}{1-\beta p^2}\right)p^{b+\frac{1}{2}}(1-\beta
p^2)^{\frac{1}{2}\left(a+\frac{1}{2}\right)-i\frac{\zeta}{2\beta}}P^{(a,b)}_n(z)\\\nonumber
=\frac{\tilde{\omega}^*C_{1;n}}{E_n+m}\sqrt{1-\beta
p^2}\left(\frac{\partial} {\partial
p}-\frac{b+\frac{1}{2}}{p}+\frac{\left(\beta\left(\frac{1}{2}-a\right)-i\zeta\right)
p}{1-\beta p^2}\right)p^{b+\frac{1}{2}}(1-\beta
p^2)^{\frac{1}{2}\left(a+\frac{1}{2}\right)-i\frac{\zeta}{2\beta}}P^{(a,b)}_n(z)\\\nonumber
=-\frac{2\beta\tilde{\omega}^*(a+n)C_{1;n}}{E_n+m}p^{b+\frac{3}{2}}(1-\beta
p^2)^{\frac{1}{2}\left(a-\frac{1}{2}\right)-i\frac{\zeta}{2\beta}}P^{(a-1,b+1)}_{n}(z)
\end{eqnarray}

In the case $k'=1-k$ that corresponds $s=-\frac{1}{2}$ and
$\xi'=\xi$ we arrive at:
\begin{equation}
R_{1;n}(p;k,\xi)=C_{1;n}p^{b+\frac{1}{2}}(1-\beta
p^2)^{\frac{1}{2}\left(a+\frac{1}{2}\right)-i\frac{\zeta}{2\beta}}P^{(a,b)}_n(z)
\end{equation}
where $a=\frac{\xi}{\beta}-\frac{1}{2}$ and $b=\frac{1}{2}-k$.

Again the relation (\ref{wf_tilde_R_2}) leads to:
\begin{eqnarray}
\tilde{R}_{2;n}(p;k,\xi)=\frac{\tilde{\omega}^*C_{1;n}}{E_n+m}\sqrt{1-\beta
p^2}\left(\frac{\partial} {\partial p}-\frac{k}{p}+\frac{\eta^*
p}{1-\beta p^2}\right)p^{b+\frac{1}{2}}(1-\beta
p^2)^{\frac{1}{2}\left(a+\frac{1}{2}\right)-i\frac{\zeta}{2\beta}}P^{(a,b)}_n(z)\\\nonumber
=\frac{\tilde{\omega}^*C_{1;n}}{E_n+m}\sqrt{1-\beta
p^2}\left(\frac{\partial} {\partial
p}-\frac{\frac{1}{2}-b}{p}+\frac{\left(\beta\left(a+\frac{1}{2}\right)-i\zeta\right)
p}{1-\beta p^2}\right)p^{b+\frac{1}{2}}(1-\beta
p^2)^{\frac{1}{2}\left(a+\frac{1}{2}\right)-i\frac{\zeta}{2\beta}}P^{(a,b)}_n(z)\\\nonumber
=\frac{2\beta\tilde{\omega}^*(b+n)C_{1;n}}{E_n+m}p^{b+\frac{3}{2}}(1-\beta
p^2)^{\frac{1}{2}\left(a+\frac{3}{2}\right)-i\frac{\zeta}{2\beta}}P^{(a+1,b-1)}_{n}(z)
\end{eqnarray}

In the end we consider the case $k'=1-k$ or equivalently as
previously $s=-\frac{1}{2}$ and $\xi'=\beta-\xi$. We arrive at:
\begin{equation}
R_{1;n}(p;k,\xi)=C_{1;n}p^{b+\frac{1}{2}}(1-\beta
p^2)^{\frac{1}{2}\left(a+\frac{1}{2}\right)-i\frac{\zeta}{2\beta}}P^{(a,b)}_n(z)
\end{equation}
where $a=\frac{1}{2}-\frac{\xi}{\beta}$ and $b=\frac{1}{2}-k$.

Having used relation (\ref{wf_tilde_R_2}) we obtain:
\begin{eqnarray}
\tilde{R}_{2;n}(p;k,\xi)=\frac{\tilde{\omega}^*C_{1;n}}{E_n+m}\sqrt{1-\beta
p^2}\left(\frac{\partial} {\partial p}-\frac{k}{p}+\frac{\eta^*
p}{1-\beta p^2}\right)p^{b+\frac{1}{2}}(1-\beta
p^2)^{\frac{1}{2}\left(a+\frac{1}{2}\right)-i\frac{\zeta}{2\beta}}P^{(a,b)}_n(z)\\\nonumber
=\frac{\tilde{\omega}^*C_{1;n}}{E_n+m}\sqrt{1-\beta
p^2}\left(\frac{\partial} {\partial
p}-\frac{\frac{1}{2}-b}{p}+\frac{\left(\beta\left(\frac{1}{2}-a\right)-i\zeta\right)
p}{1-\beta p^2}\right)p^{b+\frac{1}{2}}(1-\beta
p^2)^{\frac{1}{2}\left(a+\frac{1}{2}\right)-i\frac{\zeta}{2\beta}}P^{(a,b)}_n(z)\\\nonumber
=-\frac{2\beta\tilde{\omega}^*(n+1)C_{1;n}}{E_n+m}p^{b+\frac{3}{2}}(1-\beta
p^2)^{\frac{1}{2}\left(a-\frac{1}{2}\right)-i\frac{\zeta}{2\beta}}P^{(a-1,b-1)}_{n+1}(z)
\end{eqnarray}
It has been already noted that the last case does not have
``classical" limit when the parameters of deformation $\alpha$,
$\beta$ tend to zero. Similarly in the case of deformed algebra
with minimal length \cite{QuesneJPA05} bounded states which do not
have classical limit appear.
\section{Discussion}
In this work we considered the Dirac oscillator problem in
deformed space given by the commutation relations (\ref{algebra}).
It was shown that deformed commutation relations (\ref{algebra})
give rise to minimal uncertainty in position as well as in
momentum. To find appropriate representation for position and
momentum operators a specific nonsymplectic transformation was
proposed \cite{Mignemi_arxiv}. It allows one to find some relation
between given algebra (\ref{algebra}) and well known Snyder
algebra. Having used proposed representation it has been solved
exactly the Dirac oscillator eigenvalue problem.

It has been shown that the Dirac oscillator in deformed space with
commutation relations (\ref{algebra}) has some common features
with conventional case as well as in case of deformation with
minimal length only. A dissymmetry under the exchange of
$s=\frac{1}{2}$ with $s=-\frac{1}{2}$ that appeared in nondeformed
case due to specific substitution ${\bf P}\rightarrow{\bf
P}-im\omega{\bf X}\hat{\beta}$ takes place in case of Snyder-de
Sitter deformed algebra (\ref{algebra}). The same situation
happens in case of deformed algebra with minimal length
\cite{QuesneJPA05}. If we consider system of equations
(\ref{equ_3_b+}) and (\ref{equ_3_b-}) and make substitution
$\omega\rightarrow -\omega$ the system can be transformed to
equivalent one where $s$ is replaced by $-s$ and $E, R_1,
\tilde{R}_2$ are changed into $-E,-\tilde{R}_2, R_1$ respectively.
This transformation is valid in nondeformed case
\cite{Moshinsky_JPA89} and in the presence of deformed algebra
with minimal length \cite{QuesneJPA05}. In nondeformed situation
it is treated in connection with supersymmetry or, equivalently,
with duality between particles and antiparticles
\cite{Moshinsky_FoundPhys93}. Another similarity with previous
cases lies in the absence of negative energy $E=-m$ ground states
\cite{Moshinsky_JPA89,QuesneJPA05}.

It has been noted above the energy spectrum of the Dirac
oscillator with deformed commutation relations (\ref{algebra})
takes similar form as in case of deformed algebra with minimal
length. \cite{QuesneJPA05}. In particular, the difference
$E^2_n-m^2$ gets terms quadratic in $n$ instead of linear
dependence in nondeformed instance. It should be noted that the
relations for the energy spectrum would be in agreement with each
other if the parameter $\alpha$ in our expressions is set to zero
whereas in relations obtained in \cite{QuesneJPA05} the only
parameter corresponding to our $\beta$ is kept. We also note that
in case of deformed algebra with minimal length ground state with
energy $E^2-m^2=0$ is allowed for small values $j$ only
\cite{QuesneJPA05}. In contrast to it the Snyder-de Sitter algebra
(\ref{algebra}) does not make any restriction for parameter $j$
similarly as it was in ordinary quantum mechanics
\cite{Moshinsky_JPA89}. Ground states with nonvanishing energy
$E^2-m^2\neq 0$ are allowed for both projections of spin:
$s=\frac{1}{2}$ and $s=-\frac{1}{2}$. Here similarly to
nondeformed situation no restriction on value of total angular
momentum quantum number $j$ is imposed. It is worth stressing that
in order to have physically acceptable wavefunctions parameters of
oscillator should fulfil some conditions, namely product $m\omega$
can not take any value but it should satisfy such requirements as
(\ref{cond_gr_state=0}) or (\ref{cond_nonzero_gr_state}).

We also remark that although Dirac oscillator was introduced as a
relativistic problem in our case it is not Lorentz covariant. This
is caused by the fact that chosen algebra of operators
(\ref{algebra}) is not a relativistic one. The algebra
(\ref{algebra_2}) is obtained from the relativistic Snyder-de
Sitter algebra \cite{Mignemi_arxiv,Kowalski-glikman_PRD04} and it
seems that it easy to consider fully relativistic case but
unfortunately some problems appear. The first one is that both
time and energy will be represented by differential operators as
we have here for position and momentum operators. The second
problem is related to the behaviour of minimal uncertainties under
Lorentz transformations. These questions need careful
consideration and will examined elsewhere.

\end{document}